# The New Embedded System Design Methodology For Improving Design Process Performance

Maman Abdurohman
Informatics Faculty
Telecom Institute of Technology
Bandung, Indonesia
mma@ittelkom.ac.id

Sarwono Sutikno
STEI Faculty
Bandung Institute of Technology
Bandung, Indonesia
ssarwono@gmail.com

Kuspriyanto
STEI Faculty
Bandung Institute of Technology
Bandung, Indonesia
kuspriyanto@yahoo.com

Arif Sasongko
STEI Faculty
Bandung Institute of Technology
Bandung, Indonesia
asasongko@gmail.com

*Abstract*—Time-to-market pressure and productivity gap force vendors and researchers to improve embedded system design methodology. Current used design method, Register Transfer Level (RTL), is no longer be adequate to comply with embedded system design necessity. It needs a new methodology for facing the lack of RTL. In this paper, a new methodology of hardware embedded system modeling process is designed for improving design process performance using Transaction Level Modeling (TLM). TLM is a higher abstraction design concept model above RTL model. Parameters measured include design process time and accuracy of design. For implementing RTL model used Avalon and Wishbone buses, both are System on Chip bus. Performance improvement measured by comparing TLM and RTL model process. The experiment results show performance improvements for Avalon RTL using new design methodology are 1,03 for 3-tiers, 1,47 for 4-tiers and 1,69 for 5-tiers. Performance improvements for Wishbone RTL are 1,12 for 3-tiers, 1,17 for 4-tiers and 1,34 for 5-tiers. These results show the trend of design process improvement.

## I. INTRODUCTION

Design is an important step on whole embedded system design process. Embedded system development process begins by making hardware and software specification. The growing consumer demands for more functionality tools has lead to an increase in complexity of the final implementation of such designs. The ability of semiconductor industry to reduce the minimum feature sizes of chip has supported these demands. Also, show that the Moore's Law, roughly doubling the devices per chip every eighteen to twenty-four months, is still accurate. However, even though current IC technology is following the growing consumer demands, the effort needed in modeling, simulating, and validating such designs is adversely affected. This is because current modeling method and frameworks, hardware and software co-design environments, do not fit with the rising demands.

Fortunately, the electronic design automation industry has prepared to face this problem by providing engineers with the support for these challenging. The introduction of register transfer level (RTL) as a higher abstraction layer over gate level design is a revolution step to face this challenges. The RTL abstraction layer is accepted as the abstraction layer for describing hardware designs. The vendor of EDA is pushing the abstraction layer for addressing the lack of RTL. The definition of ESL is "a level above RTL including both hardware and software design" as suggested by The International Technology Roadmap for Semiconductors (ITRS).

ESL design and verification methodology consists of a broad spectrum of environments for describing formal and functional specifications. There are many terms used to illustrate ESL layer such as hardware and software co-design models, architectural models, RTL and software models, and cell-level models. This prescription include the modeling, simulation, validation and verification of system level designs. Models at the higher layer level are descriptions above the RTL abstraction layer depicting the system behavior. There are a number of ways to define the abstraction layer may be raised above RTL. For example, SystemC presents transaction level model for modeling and simulation of embedded software systems.

*Time-to-Market Pressure*
The growing consumer demands for various complex application led to pressure vendor to design, implement embedded system in short time frame. The late to enter market is means cost or opportunity lost.

This condition called time-to-market pressure for embedded system vendor. It needs shorter design process approach.

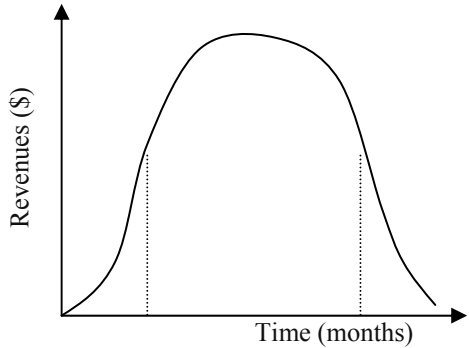

Figure 1. Time-to-Market and revenues [5]

**Embedded system design**

Design flow of embedded system begins with design specification, its define system constraint, both cost and processing time. System functionality is defined in behavioral description, hardware software partitioning is done to optimize design result and still fit the requirement. Hardware and software integration is done after hardware/software detail design. Register transfer level design is carried out by means hardware programming language such as, Verilog, VHDL and Esterel. Verification and testing process is done to ensure embedded system design is fit to specification [1].

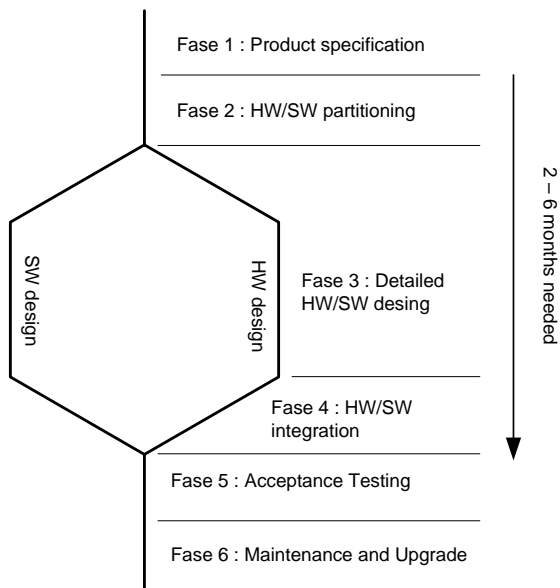

Figure 2. Embedded system design flow [1]

The embedded design process is not as simple as the concept. A considerable amount of iteration and optimization occurs within phases and between phases.

**Moore's Law and Productivity gap**

Moore's Law said that silicon capacity has been steadily doubling every 18-24 months. Its allow companies to build more complex systems on a single silicon chip. However, designer ability to develop such systems in a reasonable amount of time is not fit with the increase in complexity. This is referred to as the productivity gap, which is based on the ITRS (International Technology Roadmap for Semiconductors).

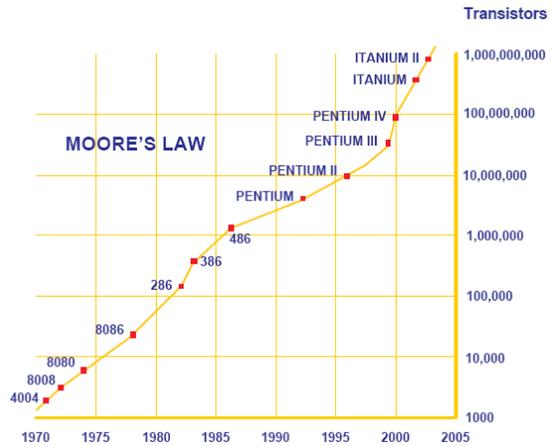

Figure 3. Moore's law [9]

Increasing the complexity and functionality of electronics systems, causes the increasing of the possible design choices and the alternatives to explore for optimization purposes. Therefore, design space exploration is vital when constructing a system in order to choose the optimal alternative with respect to performance, cost, etc. The reduction of time to develop these system-level models for optimization purposes can improve design acceleration with acceptable performance. A possible way to reduce this time is to raise the abstraction layer of design.

**Register Transfer Level design**

One of the past design revolutions in hardware design was the introduction of RTL design layer as the entry point of the design flow. At RT level, registers and a data-flow description of the transfers between them replace the gate-level instantiate of independent flip-flops and logical operators. Some hardware description languages such as VHDL, Verilog and Esterel are used for writing models at this RT level. The translation to gate level is called synthesis. Component example at this level are adder, multiplexer, decoder, and memory.

The complexity of hardware design combined with the lack of a revolution design approach similar to the RTL introduction has induced very slow simulations and caused productivity gap. The peak problem for system-on-chips is software development need, co-simulating embedded software with the RTL model is possible, but too slow to allow its effective development. Designer are forced to wait the final chip to begin writing the software of the system. This results is wasted time in the development cycle and increased time-to-market. While efficient in terms of speed, the still require the RTL model be available, they are very costly and they provide limited debugging capabilities. Another approach to face the problem is to try to raise the abstraction level : by creating models with less details before the RTL one, it should be possible to achieve better simulation speeds while at the same time less accuracy.

## II. RELATED WORK

### A. Ptolemy

Ptolemy is a project developed at the University California, Berkeley [13]. The latest Ptolemy Release is Ptolemy II 7.0.1 that has been launched u 4 April 2008. Ptolemy is a framework for simulation, prototype, and synthesis of software that has been dedicated solely to digital signal processing (DSP).

The basic concept of Ptolemy is the use of a pre-defined commutation model that will regulates inter components interactions. The main problem address by the Ptolemy is the use of the mix of various commutation models. Some of the model domains that have been implemented are: CT (continuous-time modeling), DDF(dynamic dataflow), DE (discrete-event modeling), FSM(finite state machines and modal model), PN(process networks with asynchronous message passing), Rendezvous( process), networks with synchronous message passing, SDF (synchronous dataflow), SR (synchronous reactive), Wireless.

Ptolemy II comprises supporting packages such as graphs, provides the manipulations of Graph theory, math, provides mathematical matrices and vectors and signal processing, plots, provides visual data display, data, provides type system, data wrapping and expression parses. Ptolemy II package comprises the following parts:

Ptolemy II C Code Generation: The main function is to generate codes for the SDF model, FSM and HDF: the entire model could be converted into C Codes.

Ptalon: Is an actor oriented designing representing the most commonly designing strategy in an embedded system designing. This system is frequently modeled as block diagram, where a block presents system or ;lines or inter-block arrows representing signals.

Backtracking: This facilities serve the function to save the previous system state values. The function is the most critical in a distributed computations.

Continuous domain : Continuous Domain is a remake of Continuous Time domains with meticulous semantics.

### B. COSYMA (CO-SYnthesis for eMbedded Architectures).

The Cosyma is developed by the Braunschweig University/ The Cosyma performs operation-separation process on the lowest blocks to improve the speed of program execution time.

This speed improvement is achieved by adding co-processors hard ware that will perform part of the functions that traditionally run by software. The following Figure indicates Cosyma flow diagram. Its inputs comprises the Cx program (8). It is an extension of the C Program to enhance parallel data processing. Its final output is hardware block and the primitive of communication in hardware software.

### C. VULCAN

Vulcan is designed to cut cost of ASIC Program Time. The cost reduction could be achieved by separating designing parts into software.

Initial specification is written in Hardware-V, i.e., the HS-Description Language (HDL_. That could be synthesis trough OLYMPUS synthesis System. Specifications in C would be mapped into representations between Control-Data Flow Graph (CDFG). It is this level that Vulcan separates hardware from software.

The separation of hardware from software is achieved through heuristic graphs partition algorithm that work under polynomial time. This separation algorithm has paid its attention on different partitions on CDFG between Hardware and software, to minimize hardware costs but simultaneously maintain the predetermined deadlines.

TABLE 1. COMPARISON OF MODELING FRAMEWORKS

| Name | Specification | Modeling | HW/SW Part | SW design | HW design |
|---|---|---|---|---|---|
| Ptolemy | C++ | FSM | GCLP | C | VHDL |
| Cosyma | C* | Syntax DAG | Sim annealing | C | HardwareC |
| Vulcan | Hercules | Vulcan | Greedy | DLXCC | HEBE |
| Stellar | Nova | Nebula | Magellan | GCC | Asserta |

### D. STELLAR

STELLAR is a system level co-synthesis environment for transformative application. A transformative transformation is an application that executes processes every time it has a trigger such as JPEG Encoder. As an input specification, STELLAR provides a C++ Library with ist NOVA name.
The inputted specifications are in forms of application specifications, the architecture and performance yardsticks. The outer format of the NOVA is executable. STELLAR supports software estimation through profiling and using ASSERTA synthesis device in estimating hardware.

STELLAR get input specification and definitions in NEBULA mediating format. Its designing environment provides two devices: MAGELLAN and ULYSSES of co-synthesizing and evaluation the GW-SW. the MAGELLAN optimize latency retimes and the ULYSSES OPTIMIZES the applications throughput. Its outer part comprises hardware specification, software and interface. The exterior specification could be translated into SystemC code. And its functionalities would be verified through simulation.

Table 1 shows the comparison between all embedded system design frameworks.

## III. THE NEW DESIGN METHODOLOGY

### A. Transaction Level Modeling (TLM)

Transaction-level Modeling fills the gap between purely functional descriptions and RTL model. They are crated after hardware/software partitioning, that is, after is has been decided which, for each processing, if it would be done using a specific hardware block or by software. The main application of TLM is to serve as a virtual chip (or virtual platform) on which the embedded software can be run.

The main idea of TLM is to abstract away communication on the buses by so-called transactions : instead of modeling all the bus wires and their state change, only the logical operations (reading, writing etc) carried out by the busses are considered in the model. In contrary to the RTL, where everything is synchronized on one or more clocks (synchronous description), TLM models do not use clocks. They are asynchronous by nature, with synchronzation occuring during the communication between components. These abstractions allow simulations multiple orders of magnitude faster than RTL.

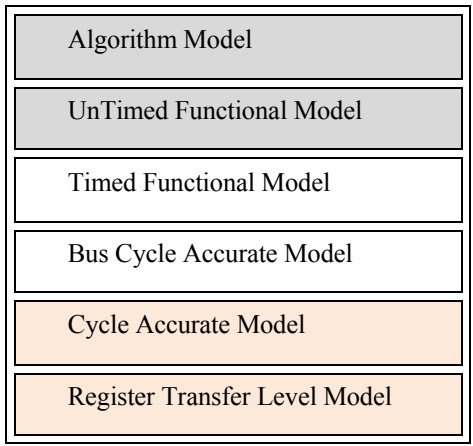

Figure 4. TLM Model Stack

The other advantage of TLM models is that they require far less modeling effort than RTL or than Cycle Accurate model. This modeling effort is further reduced when there alreay exists a C/C++ functional code for the processing done by the hardware block to model. For instance, one can reuse the reference code for a video decoder or for a digital signal processing chain to produce a TL model. Unlike a Cycle Accurate model, which is no longer the reference after RTL is created, TLM is by this means an executable, "golden model" for the hardware. Various definitions of TLM exist; some of them even rely on clocks for synchronization, which looks more like Cycle Accurate level. A transaction term is an atomic data exchange between an initiator and target. The initiator has the initiative to do the transaction whereas the target is considered as always able to receive it (at least, to indicate to the initiator that it is busy). This corresponds to classical concepts in bus protocols. The initiator issues transactions through an initiator port, respectively a target receives them by a target port. Some components only have initiator ports some have only targets ports. Also, some components contain both initiator and target ports.

The information exchanged via a transaction depends on the bus protocol. However, some of them are generally common to all protocols :

- The *type* of transaction determinates the direction of the data exchange, it is generally read or write.
- The *address* is an integer determining the target component and the register or internal component memory address.
- The *data* that is sent to received.
- Some additional *meta-data* including : a return status (error, success, etc), duration of the transaction, bus attributes (priority, etc).

The most basic functionality shared by all buses or more generally interconnection networks is to route the transactions to their destination depending on their address. The destination is determined by the global memory address map which associates a memory range to each target port.

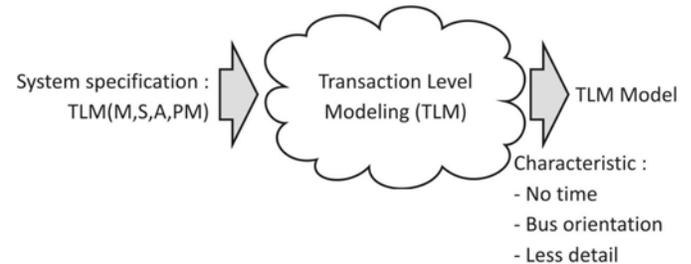

Figure 5. TLM process model

In order for the embedded software to execute correctly, the address map, the offset for each register must be the same as in the final chip (register accuracy). Additionally, the data produced and exchanged by the components must also be the same (data accuracy). Finally, the interrupts have to correspond logically to the final ones. One can view these requirements as a contract between the embedded software and the hardware. This contract guarantees that if the embedded software runs flawlessly on the virtual platform, then it will run in the same way on the final chip.

*B. New Design Flow of Hardware Embedded System Design*

In this paper, the new design flow for modeling hardware embedded system is designed using transaction level modeling (TLM) method for early verification purpose. Verification process done at the first step before detail design. Transaction level modeling is one of new trends on embedded system design after the development of register transfer level modeling.

The research scope is particularly on hardware embedded system after performing separation of hardware and software process. There are three stages in detailed design:

1. Hardware part definition: hardware embedded system definition that will be implemented.

2. TLM modeling: Model construction with transaction modeling approach and perform early verification. Model refinement process can be generate by performing tuple correction : M, S, A, PM.

3. RTL modeling: RTL model construction is the final process of all hardware designs of embedded system. In

this process, transformation from TLM model into RTL model is conducted.

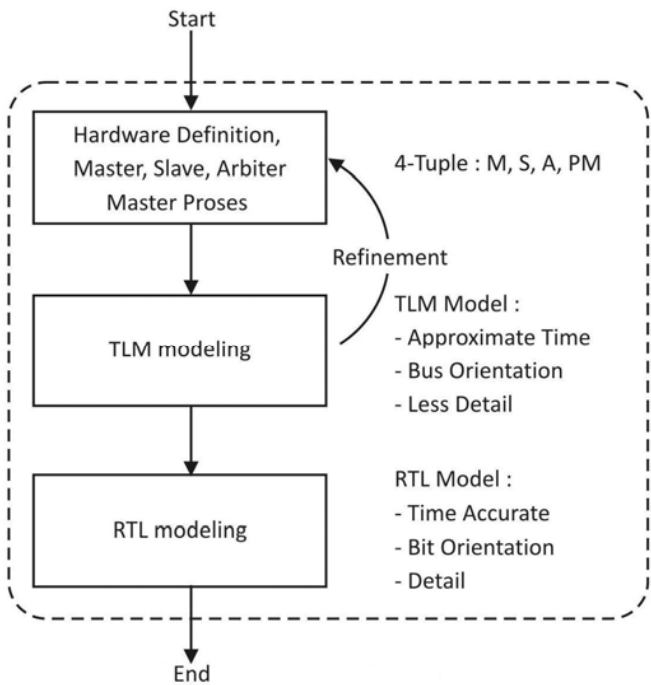

Figure 6. The new design methodology

### C. Procudure and Modeling Diagram Block

Basic procedure of modeling is designed as standard process on hardware modeling. Modeling steps of new design methodology are:

1. Define : 4-tuple input (M, S, A, PM).
2. A module with port and method is made for each master.
3. A module with port and method is made for each slave.
4. An arbiter bus is made with algorithm in A.
5. Every method in master and slave is defined in PM.
6. Early verification of system requirement compliance
7. If system requirement is not satisfying, then perform tuple refinement starting from step 1.
8. Adding port and RTL process
9. Port and process removal from TLM.
10. RTL arbitrary implementation.
11. Port mapping

Stages 1 to 6 are initial stage of transaction level modeling creation in the purpose of early verification of hardware modeling. The first process output is a TLM model that fulfill the design requirements.

### 1) Diagram Block

Diagram block is a diagram that shows many inputs and outputs of the system. Inputs of diagram block of transaction level modeling include :

- Master : Number of master component actively perform read () and write process as standard operation of components
- Slave : Number of slave components considered passive components and waiting for transaction of master.
- Arbiter : bus management system, namely mutual access management algorithm of one slave with one master or more.
- PM : Process taking a place in master and slave such as read() and write() process.
- Tiers : Total the whole main components existing in a system including Arbiter.
- Specification is system requirement explanation that should be met by system designed.

Output of design system block is a TLM Model. Model formulation process is conducted systematically.

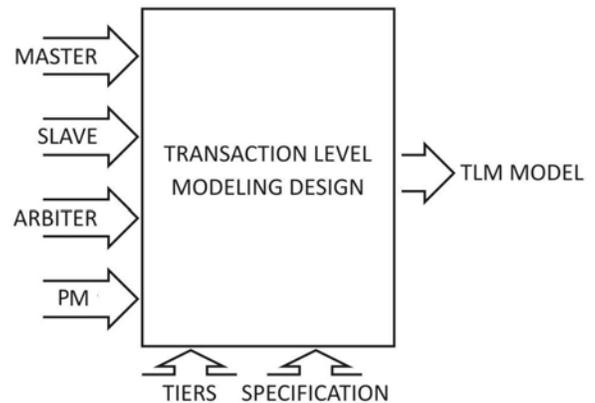

Figure 7. Diagram Block

### 2) Defining Master and Slave

- Master and slave component definition consists of three parts; name, port, and functions/method. Example of master:

| Name : MicroPro |
|---|
| Port :<br>  int mydata, input1, input2, input3;<br>  int cnt = 0;<br>  unsigned int addr = m_start_address; |
| Function/Method :<br>Do_add();<br>Memory_Read();<br>Memory_write(); |

- Arbiter is bus management algorithm, such as: round robin.
- PM is a process in master. PM is the more detail definition of in the form of pseudo code.

In transaction level modeling, data transfer process and control from master to slave are conducted by accessing a bus controlled by an arbiter. Each master can deliver request of bus access to send data or read data from slave. There will be several possible conditions achieved by master; they are bus condition is OK if bus is not being operated by other masters or WAIT condition if bus is being used by other master or ERROR condition if targeted slave is not around in slave list.

*3) TLM – RTL Transformation*

After finishing early verification process and being met with given specification, then the last stage is transformation from TLM into RTL. The purpose of the transformation is to generate detail model available for synthesis. Phases of TLM into RTL model transformation can be divided into several general stages; those are:

- Port addition and deletion: in the process of TLM modeling, there should be ports that are required to delete, because basic principle is not needed in RTL model, such port request. Meanwhile, it is necessary to add new ports in RTL model for performing detail process, as the nature of RTL modeling.
- Process addition and deletion: In spite of ports addition and deletion, it is also necessary to add and delete process. Example of process that must be deleted from RLM is such process that tries to send request, while addition process that should be given in RTL model is process of accessing multiplexer.
- Total Master and Slave determination: Total master and slave is used to make pattern of RTL bus. Total master and slave can influence total multiplexers and types of multiplexer. Multiplexer for 4 masters applies the first mux4 while 2 masters apply the first mux2.
- Determining arbitrary (according to given protocol) Arbitrary is management algorithm of slave access when the access is from one master or more. Example of algorithm used is round robin, such as in Avalon bus.
- Port mapping : The last stage of transformation is connecting all ports from all components available along with additional components, such as multiplexer, detail, pin-per-pin.

*4) Examples of TLM-RTL Transformation*

The followings are examples of transformation from TLM to RTL by using RTL bus with Wishbone.

Bus target: **Wishbone**

1. Port addition and deletion:

- Sc_in_clk clock; (added)
- Sc_port<sc_signal_out_if<bool>> grantM1; (deleted)

2. Process addition and deletion:

- Sel_mux_master1 (added)
- grantM1process (deleted)

3. Total Master and Slave determination:

- Determining the amount of rows added and reduced of all systems.

4. Arbitrary determination

- Wishbone protocol arbitrary is Round robin
- Every master sends request of slave access. If there are several masters requesting access of one similar slave, then the arbiter will give an access for the master and send waiting signal for other masters.

5. Port mapping of all modules: master and slave

- Mapping of master post to all multiplexers.
- Mapping of multiplexer post to slave and the master.
- Mapping of multiplexer post to slave and the master.

*D. Criteria and Measurement*

There are two criteria used for measuring new system experiment, they are :

1. Design performance improvement (Te)

Performance improvement is characterized by the decrease of time required to design embedded system. Design period by using new method will be compared with RTL design period. Time difference needed to design the same systems from two different methods will be considered the success of new design. New design system is considered success if design period needed is shorter than the previous time design. General formula of design performance improvement (Te) is Te = $T_{RTL}/T_{TLM}$. $T_{RTL}$ is design process time for modeling RTL model and $T_{TLM}$ is design process time for modeling TLM.

2. Target of Criteria: Accuracy level ($\alpha$)

Design model difference can be conducted in the purpose of improving performance. This can be accepted if both models can bring about the same output for the same input. The closer the result of both systems to the same input is, the more accurate the system is. Accuracy level ($\alpha$) = P(Input) – P(Input) $\approx 0$.

IV. EXPERIMENT AND RESULTS ANALYSIS

*A. Avalon and Wishbone Bus for on Chip System (SoC)*

Avalon and Wishbone bus are busses for SoC. The bus is designed for chip-based application. SoC is a compact system with three kind components, master and slave components and bus system.

Avalon and Wishbone buses are used in implementation stage in the level of RTL. There are 5 main components in Avalon bus along with each function as follows:

1. Master : active components which have initiative to perform data access either read() or write().

2. Slave : passive component waiting for data access from master.

3. Logic Request : components managing access requests from master for slave. each component has one logic request component.

4. Logic Arbitrator: component managing access of one slave according to request of one master of more. Each slave has one Logic Arbitrator to manage the slave access.

5. Multiplexer : component for managing access of a slave according to request of Logic Arbitrator. There are 5 multiplexers for each slave; mux address, mux BE_n, mux write, mux writedata and mux read. There is one mux for data displayed for master; mux master.

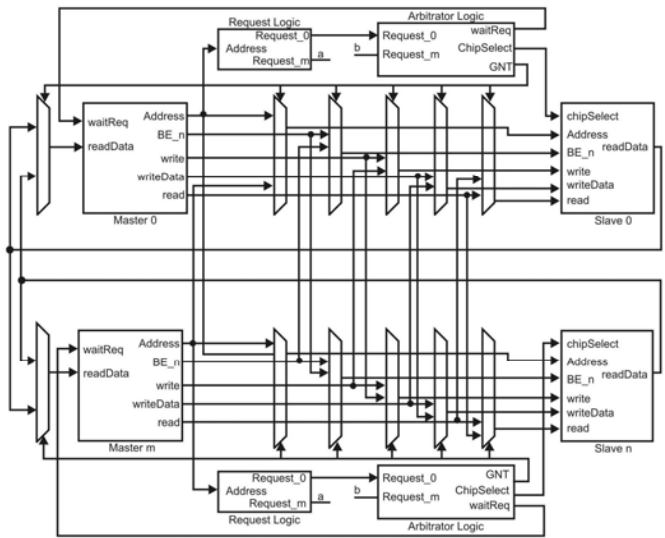

Figure 8. Avalon bus architecture

There are 5 main components in Wishbone bus include : master, slave, decoder, round robin arbiter and multiplexor. The function of each component the same as on Avalon bus.

### B. Testing Scenario

Test is performed to measure performance increase of design by using new design flow of transaction level modeling (TLM – Transaction Level Modeling) compared to Level Register Transfer modeling (RTL – Register Transfer Level). Some testing scenario will be conducted in testing process involving several master components, slave and arbiter.

On each testing scenario, testing model is generated in transaction level modeling and RTL. Both of the models will be compared based on the line amount required to implement. Testing is conducted starting from simple system, consisting of a master and a slave. Then, testing component complexity will be ignored periodically by adding the amount of tiers continuously.

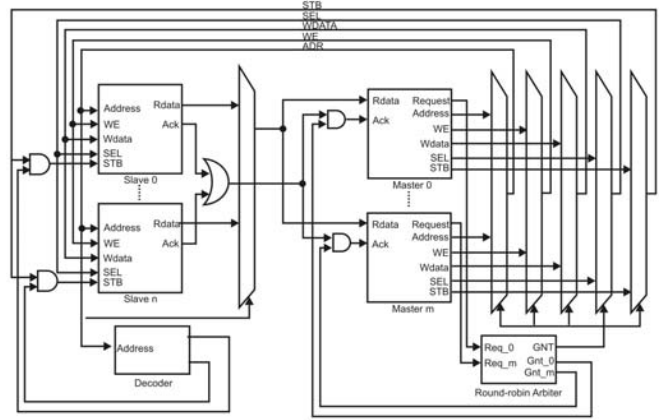

Figure 9. Wishbone bus architecture

Tiers are general terms of displaying embedded system components which communicate each other, for example, 2-tiers means there are two components communicating each others, 3-tiers means there are 3 components communicating each others, and so on.

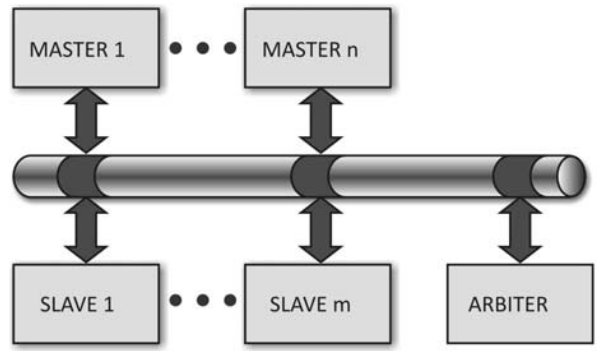

Figure 10. multi master-slave system

### C. TLM-RTL Model Testing

#### a. Line of Code of TLM-RTL Model

The testing involves two, three, four and five components, those are master, slave and arbiter. Master component actively generates and sends data to slave, while slave serves as receiver.

Based on the experiment results, it suggests that the amount of lines needed to model system by using TLM is less than that of using RTL. Such condition can take a place because master, slave, and bus definition on RTL is more detail than that on TLM.

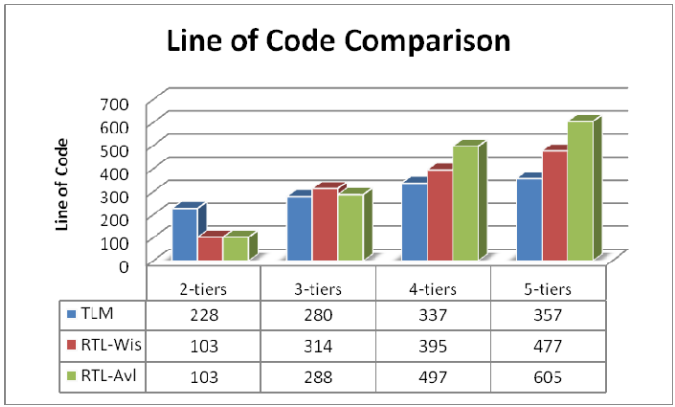

Figure 11. Line of Code comparison

Detail level of modeling on RTL can influence several parts of program, including:

- Port definition of each master and slave component.
- Initial definition of top level system, including port addition, instantaneousness, port mapping and destruction.
- New component definition called as multiplexor. On Avalon bus of each slave addition, 6 new multiplexor shall be added accordingly.

In TLM, component definition process and port mapping is simpler than that of in RTL, so it does not require many instructions as compared to RTL. New port mapping when master addition takes a place is a mapping with bus and clock.

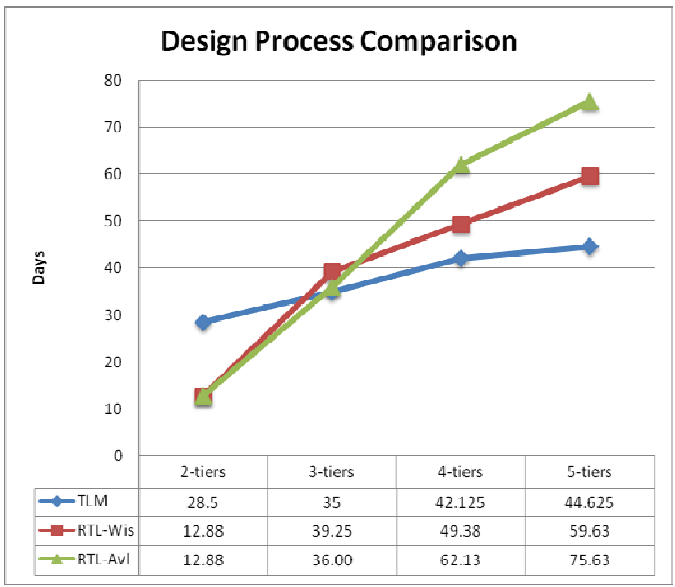

Figure 12. Design process time comparison

**b. Measurement is based on design time (man days).**

Design time refers to time needed by programming for code generation and report. Standardarization used is 8 line codes per man day. Design time can be directly decreased from the total modeling code for each TLM and RTL modeling. Figure 12 shows comparison result of time needed to design the four testing scenario.

Measurement of Design Process Performance (Te)

Performance is one of the important parameters to measure the success of new method. In this dissertation, design process performance can be measured according to the comparison between design process times needed by using RTL model compared with TLM model.

The measurement of performance improvement of design process can be conducted by using the following equation:

$$T(e) = Trtl / Ttlm$$

Based on the experiment result conducted as shown in Figures 11 and 12, it can be concluded that performance improvement graph as shown in Figure 13 can be obtained.

As shown in Figure 13 it indicates that design process performance increases as the increase of amount of component on the embedded system except for case study of 2-tiers whose performance is higher than that of 3-tiers.

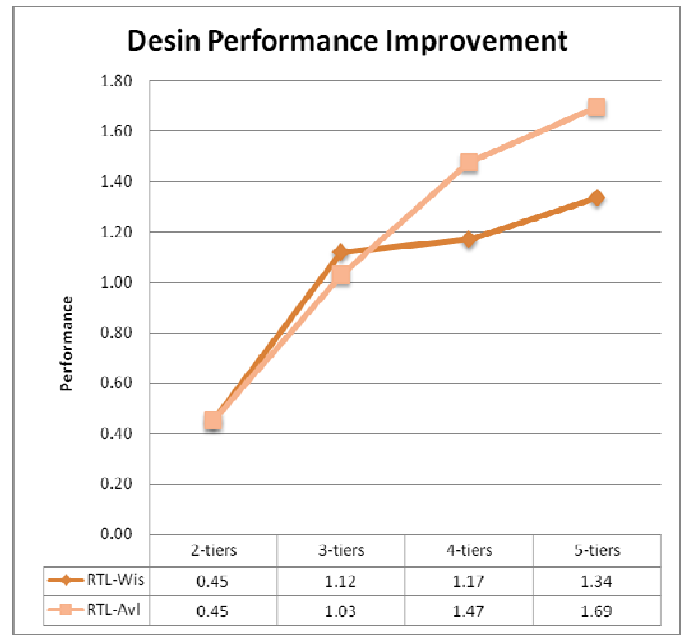

Figure 13. Design performance improvement

One of the advantages of TLM modeling is that transaction will occur among components. The more components the system is, the higher increase of transaction level by the use of bus is. Transaction improvement of components is very appropriate to TLM modeling. In RTL level modeling, in the contrary, the amount of components and transactions of a system will make the design difficulty higher.

Therefore, total components and transaction in the modeling are getting bigger and bigger, and thereby design process will get slower. Based on the condition mentioned above, design process on TLM modeling is getting better and better under the circumstances that there are many components making interaction each others. Those are the advantages of TLM level modeling taking a place in transaction level.

V. CONCLUSION

Based on the testing shown in the previous chapter, it can be concluded that there are several important things, including:

1. The new embedded system design flow can be used to increase design process performance. It means that using this method the design process will shorter than RTL modeling with performance improvement compare to RTL Avalon bus are 1.03, 1.47, 1.69 for 3,4 and 5 tiers respectively. The performance improvement compare to RTL Wishbone bus are 1.12, 1.17 and 1.34 for 3,4 and 5 tiers respectively.

2. TLM level modeling will be better implemented in a complex system, in the condition of more than two components having interactions in which there occurs arbitrary process.

Contribution of this paper are the new embedded system design flow by using transaction level modeling approach and standard procedure to design hardware RTL model.

In the effort of constructing an integrated framework ranging from specification till model construction that is available to be synthesized, the following discussion shall be conducted in the next research:

1. Automation process of all standard procedure
2. Design of parts of system software and application
3. Integrated process between software and hardware of software.

By adding the three parts of processes, then new framework will soon be generated under embedded system design. The initial framework includes systematic steps of embedded system construction. The next process is automation of the whole processes.

ACKNOWLEDGMENT

Maman Abdurohman thanks to the Faculty of Informatic IT Telkom and Faculty of STEI Electro Bandung Institut of Technology for their financial support and research resources so that this research could be completed.

AUTHORS PROFILE

Maman Abdurohman is a PhD student at STEI faculty of Bandung Institute of Technology. He is working at Informatics faculty of Telecom Institute of Techonolgy – Bandung. His primary areas of interest include embedded system design and microcontroller. Maman has an master degree from Bandung Institute of Technology. Contact him at mma@ittelkom.ac.id

Kuspriyanto is a Professor at STEI faculty of Bandung Institute of Technology. He is a senior lecturer in computer engineering laboratory. His major areas of interest include digital system and electronic design. His job is Head of Laboratory of Computer Engineering. Contact him at kuspriyanto@yahoo.com

Sarwono Sutikno is a Associate Profesor at STEI faculty of Bandung Institute of Technology. His major areas of interest include cryptography and embedded system design. His job is a director in PPATK on electronic transaction control. Contact him at ssarwono@gmail.com

Arif Sasongko is a lecturer at STEI faculty of Bandung Institute of Technology. His major areas of interest include wimax design and embedded system design. His current project is designing highspeed data link wimax system. Contact him at asasongko@gmail.com